\documentclass[a4paper,11pt]{article}
\pdfoutput=1 

\usepackage{jinstpub} 
\usepackage{siunitx}
\usepackage{graphicx}
\usepackage{subfig}
\usepackage{appendix}
\usepackage{wasysym}
\usepackage[nopostdot,nogroupskip,style=super,nonumberlist,toc,automake,toc]{glossaries} 

\usepackage{lineno}

\makeglossaries

\newglossaryentry{TiPC}{name={TiPC},description={Timed Photon Counter}}
\newglossaryentry{PMT}{name={PMT},description={photomultiplier tube}}
\newglossaryentry{SE}{name={SE},description={secondary electron}}
\newglossaryentry{PE}{name={PE},description={primary electron}}
\newglossaryentry{TSE}{name={TSE},description={transmission secondary electron}}
\newglossaryentry{FSE}{name={FSE},description={forward-scattered electron}}
\newglossaryentry{RSE}{name={RSE},description={reflection secondary electron}}
\newglossaryentry{BSE}{name={BSE},description={backscattered electron}}
\newglossaryentry{TEY}{name={TEY},description={transmission electron yield}}
\newglossaryentry{REY}{name={REY},description={reflection electron yield}}
\newglossaryentry{ALD}{name={ALD},description={atomic layer deposition}}
\newglossaryentry{DRIE}{name={DRIE},description={deep-reactive ion etching}}
\newglossaryentry{PR}{name={PR},description={photoresist}}
\newglossaryentry{XPS}{name={XPS},description={X-ray photoelectron spectroscopy}}
\newglossaryentry{PECVD}{name={PECVD},description={plasma-enhanced chemical vapor deposition}}
\newglossaryentry{LPCVD}{name={LPCVD},description={low-pressure chemical vapor deposition}}
\newglossaryentry{SEM}{name={SEM},description={scanning electron microscope}}
\newglossaryentry{tynode}{name={tynode},description={transmission dynode}}
\newglossaryentry{MEMS}{name={MEMS},description={micro-electro-mechanical systems}}

\title{\boldmath The construction and characterization of MgO transmission dynodes}

\author[a,b,1]{H.W. Chan\note{Corresponding author.}}
\author[a,b]{V. Prodanovi\'c}
\author[c]{A.M.M.G. Theulings}
\author[d]{S. Tao}
\author[e]{J. Smedley}
\author[c]{C.W. Hagen}
\author[b]{P.M. Sarro}
\author[a]{and H. v.d. Graaf}

\affiliation[a]{National Institute for Subatomic Physics (NIKHEF),\\Science Park 105, 1098 XG, Amsterdam, The Netherlands}
\affiliation[b]{Faculty of Electrical Engineering, Mathematics, and Computer science, Department of microelectronics/ECTM, Delft University of Technology\\Feldmannweg 17, 2628 CT, Delft, The Netherlands}
\affiliation[c]{Faculty of applied sciences, Department of Imaging Physics, Delft University of Technology,\\Lorentzweg 1. 2628 CJ, Delft, The Netherlands}
\affiliation[d]{Materials Simulation and Modelling, Department of Applied Physics, Eindhoven University of Technology, 5600 MB, Eindhoven, The Netherlands}
\affiliation[e]{SLAC National Accelerator Laboratory, Stanford University, 2575 Sand Hill Road, Menlo Park, California 94025, USA}

\emailAdd{h.w.chan@hotmail.com}

\abstract{In this work we demonstrate that ultra-thin (5 and \SI{15}{\nm}) MgO transmission dynodes with sufficient high transmission electron yield (\gls{TEY}) can be constructed. These transmission dynodes act as electron amplification stages in a novel vacuum electron multiplier: the Timed Photon Counter. The ultra-thin membranes with a diameter of \SI{30}{\um} are arranged in a square 64-by-64-array. The TEY was determined with a scanning electron microscope using primary electrons with primary energies of 0.75 - \SI{5}{\keV}. The method allows a TEY map of the surface to be made while simultaneously imaging the surface. The TEY of individual membranes can be extracted from the TEY map.  An averaged maximum TEY of $4.6\pm0.2$ was achieved by using \SI{1.35}{\keV} primary electrons on a TiN/MgO bi-layer membrane with a layer thickness of 2 and \SI{5}{\nm}, respectively. The TiN/MgO membrane with a layer thickness of 2 and \SI{15}{\nm}, respectively, has a maximum TEY of $3.3\pm0.1$ (\SI{2.35}{\keV}). Furthermore, the effect of the electric field strength on transmission (secondary) electron emission was investigated by placing the emission surface of a transmission dynode in close proximity to a planar collector. By increasing the electric potential between the transmission dynode and the collector, from -\SI{50}{\V} to -\SI{100}{\V}, the averaged maximum TEY improved from $4.6\pm0.2$ to $5.0\pm0.3$ at a primary energy of \SI{1.35}{\keV} with an upper limit of 5.5 on one of the membranes.} 


\arxivnumber{0000.0000} 

\begin{document}
\maketitle
\flushbottom


\section{Introduction}
\label{sec:intro}

Vacuum electron multipliers, such as photomultiplier tubes (\gls{PMT}s), are versatile photon detectors with high gain and low noise \cite{Hamamatsu2007}. They are essential in single photon counting applications, which can be found in, among others, high energy physics experiments and medical imaging. The operating principle of a PMT is the conversion of photons into photoelectrons and subsequent electron multiplication in vacuum. The sensitivity of the photocathode depends on the photocathode material and can be tailored to a part of the spectrum ranging from infrared to ultraviolet. An incoming (photo-)electron that impacts on the surface of a reflective dynode generates multiple secondary electrons (\gls{SE}s). An intricate reflective geometry is required to ensure that the SEs are directed from dynode to dynode and eventually be collected by an anode. A disadvantage of this design is that PMTs are sensitive to external magnetic fields which can disturb the electron paths within the device. In addition, the detectors have low spatial resolution due to their bulkiness and are expensive to fabricate. 

\begin{figure}
\centering
\includegraphics[width=0.8\textwidth,origin=c,angle=0]{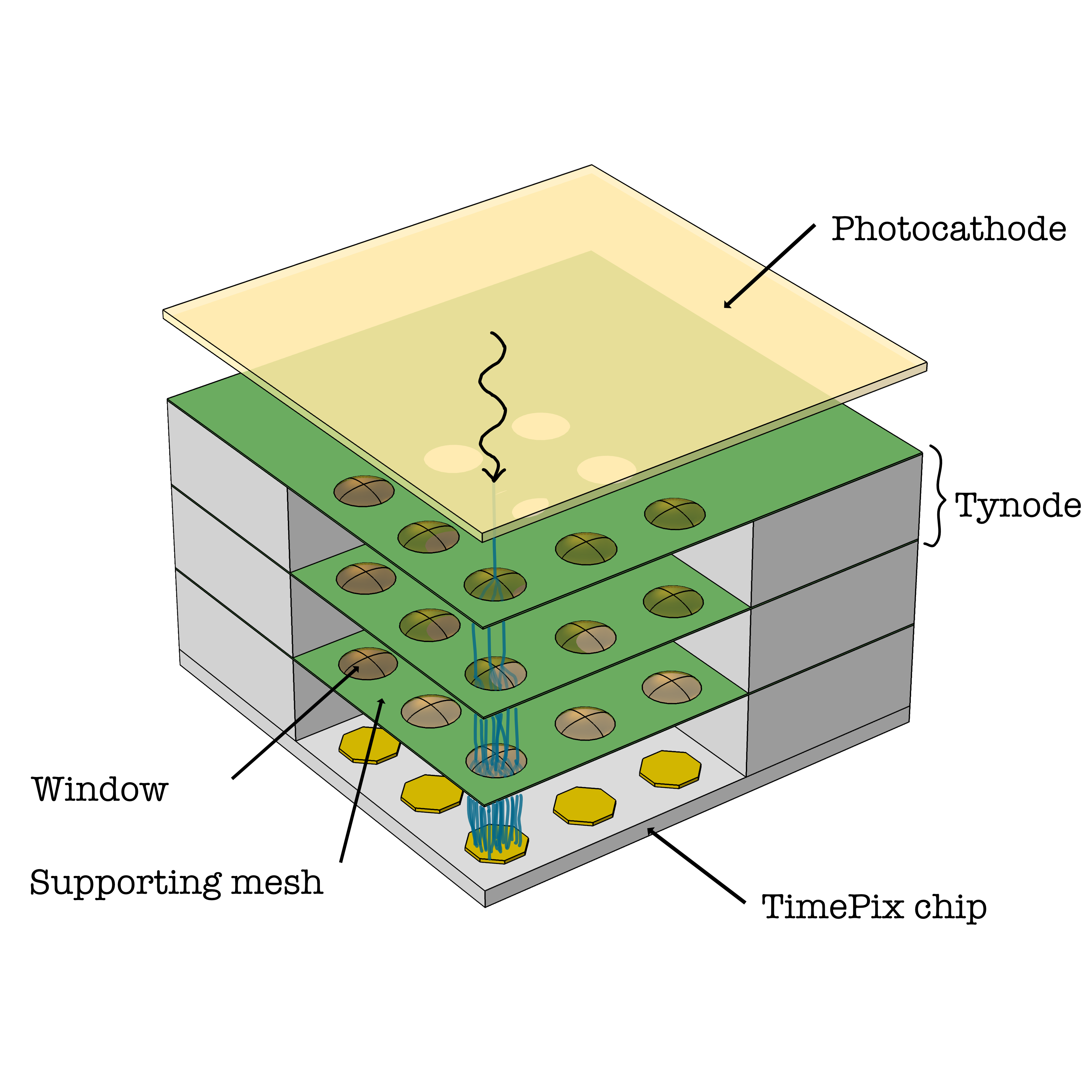}
\caption{\label{fig:1} The Timed Photon Counter is a novel vacuum electron multiplier that consists of a photocathode, a tynode stack and a TimePix chip in an enclosed vacuum package \cite{VanderGraaf2017}. The ultra-thin membranes are suspended in a supporting mesh and are spaced to align with the pixel pads of a TimePix chip. A photon that hits the photocathode is converted to a photoelectron. The photocathode is at a negative potential with respect to the first tynode. As a consequence, the photoelectron accelerates towards the first tynode and gains energy equal to $E=q\Delta V$, where $q$ is the charge of the electron and $\Delta V$ the potential difference. When the photoelectron impacts the top of the first tynode, multiple SEs are emitted from the bottom. Subsequently, the SEs are accelerated and gain energy before they impact the next tynode. The process repeats until the avalanche of electrons is collected by the pixel pad of the TimePix chip. For a stack with $m$ tynodes, the gain is $G=\sigma_T^m$, where $\sigma_T$ is the transmission electron yield per tynode.}
\end{figure}

The Timed Photon Counter (\gls{TiPC}) is a novel electron multiplier that utilizes transmission dynodes (\gls{tynode}s) for electron multiplication \cite{VanderGraaf2017}. The mode of operation of a transmissive vacuum electron multiplier allows for a compact, planar and closely-stacked design (Figure \ref{fig:1}), which outperforms traditional reflective electron multipliers, such as PMTs, in terms of temporal and spatial resolution. The electric field between the multiplication stages are stronger and more homogeneous in comparison to the electric fields in reflective geometries used in PMTs. As a result, the transit time of the SEs will be only tens of picosecond with a small spread, which results in an improved temporal resolution. An additional benefit of the stronger electric field is the reduced susceptibility to external magnetic fields that might disrupt the detector.

The core innovation in TiPC are the ultra-thin transmission dynodes. A primary electron (\gls{PE}) with sufficient energy can penetrate a thin membrane. The range of a PE in bulk material is given by $R=CE_0^n$, where $E_0$ is the PE energy, $C$ is a material-dependent constant and $n$ is a constant that depends on the energy of the PE \cite{Kanaya1972, Fitting1974}. In the process, the PE loses energy and generates internal SEs along its track. Internal SEs near the surface of the membrane have a chance to escape into vacuum. In the case of a thin membrane, this can either be on the side the PE entered or the opposite side, resulting in reflection and transmission SEs, respectively. 

\begin{figure}
\centering
\includegraphics[width=0.6\textwidth,origin=c,angle=0]{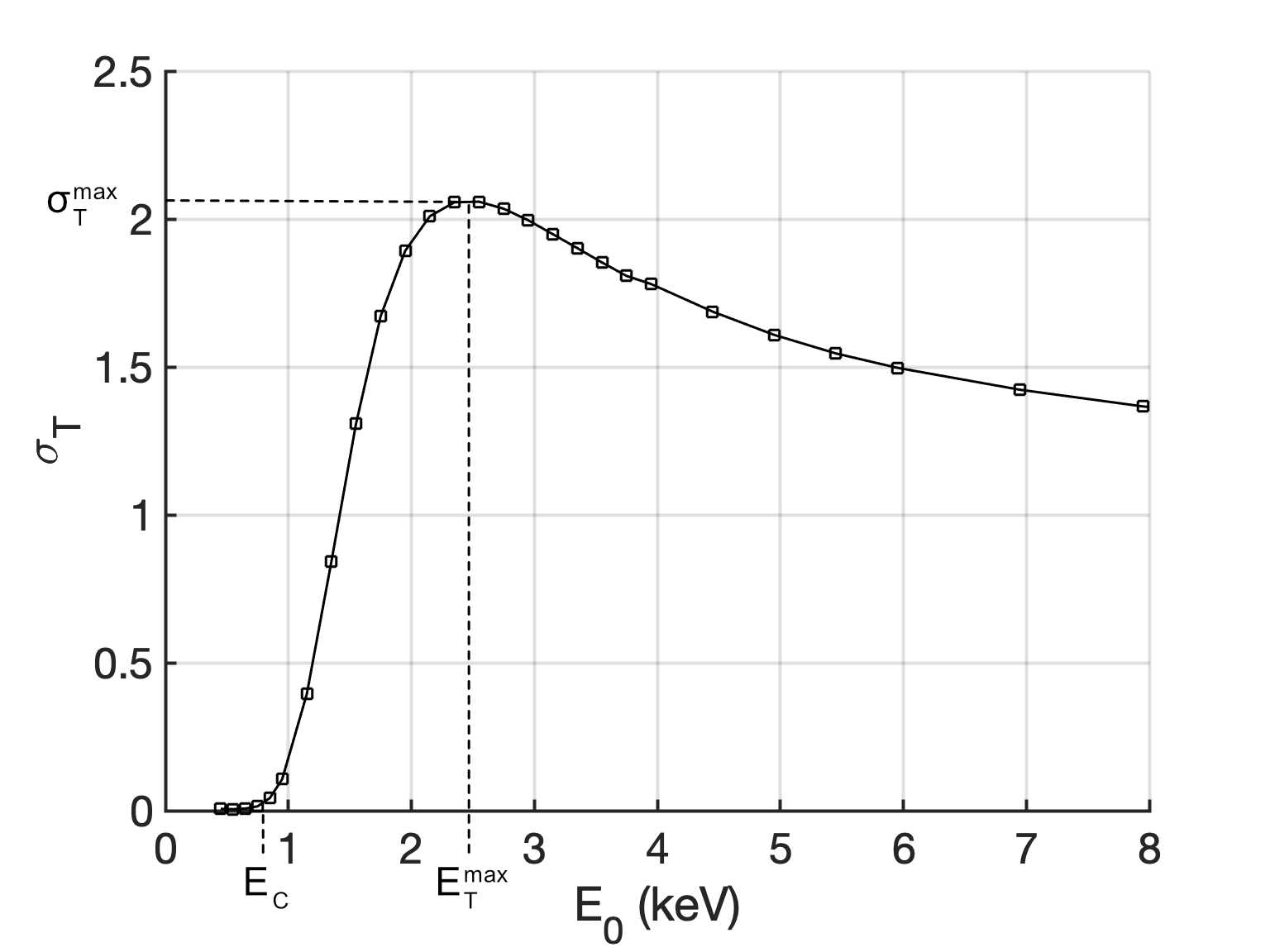}
\caption{\label{fig:2} A typical TEY curve with two distinctive characteristics that depend on film thickness: the critical energy $E_c$ and the maximum TEY $\sigma_T^{\text{max}}(E_T^\text{max})$. The performance of a tynode is expressed by the latter. This TEY curve was determined for a TiN/Al\textsubscript{2}O\textsubscript{3} membrane with layer thicknesses of 5.7/\SI{25}{\nm} \cite{Chan2021}.}
\end{figure}

For TiPC, we aim to fabricate tynodes with a transmission electron yield (\gls{TEY}) of 4 or higher for 1-keV-electrons \cite{VanderGraaf2017}. The gain of the detector is given by $G=\sigma_T^m$, where $\sigma_T$ is the TEY per tynode and $m$ is the number of multiplication stages. A design with 5 tynodes with $\sigma_T=4$ will have a gain of 1024, which is above the detection threshold of the TimePix pixels \cite{Llopart2007}. A tynode is characterized by its energy-yield-curve (see figure \ref{fig:2}), which is determined by measuring the TEY while incrementally increasing the PE energy $E_0$. The yield curve $\sigma_T(E_0)$ has two features that depend on the thickness of the tynode: the critical energy $E_c$ and the maximum TEY $\sigma_T^{\text{max}}(E_T^\text{max})$. The former is defined as the threshold where $1\%$ of the PEs crosses the entire membrane. It coincides with the onset of the yield curve. The latter is used to express the performance of a tynode, since it contains the two benchmarks that concerns TiPC: the maximum TEY and PE energy. 

Although one of the first working transmission dynodes were made in the 60's by Sternglass and Wachtel \cite{Sternglass1956}, the thickness of the transmission dynode was one of the limiting factors for wider application. In their case, the optimum film thicknesses for a KCl/Au/SiO film was found to be 60/2/\SI{10}{\nm} respectively, which have a maximum TEY of 8.4 at \SI{3.2}{\keV}. A review on a variety of thin film materials has shown that high yields can be achieved for alkali halides, semiconductors and insulators, though the required PE energy is often a few to tens of keVs \cite{Tao2016}. The thicknesses of these membranes are usually above \SI{100}{\nm} due to the complexity of fabricating freestanding thin films.   

As part of the MEMBrane project, we aim to fabricate ultra-thin tynodes using micro-electro-mechanical system (\gls{MEMS}) fabrication techniques. The advancement in MEMS technology allows for the creation of ultra-thin membranes. For instance, electron transparent windows with a thickness of \SI{10}{\nm} are used in the design of MEMS nanoreactors \cite{Creemer2010}. The silicon nitride film was deposited by low-pressure chemical vapor deposition (LPCVD) and released by subsequent chemical and plasma etching. Using a similar process, we have constructed a tynode with LPCVD SiN membranes \cite{Prodanovic2018}. The ultra-thin membranes with a thickness of $\SI{40}{\nm}$ and a diameter of \SI{30}{\um} were arranged in a 64-by-64 array (see figure \ref{fig:1}) and have a TEY of 1.6 ($\SI{2.85}{\keV}$). In addition, a tynode with a different membrane material was fabricated by means of atomic-layer-deposition (\gls{ALD}). A TEY of 2.6 ($\SI{1.45}{\keV}$) was measured on an ALD Al\textsubscript{2}O\textsubscript{3} membrane with a thickness of $\SI{10}{\nm}$. On top of the membrane, a titanium nitride (TiN) layer with a thickness of \SI{2}{\nm} was sputtered, which provides lateral conductivity. In another design, the TiN layer was encapsulated between two Al\textsubscript{2}O\textsubscript{3} layers, which improved the reliability of the conductive layer \cite{Chan2021}. A TEY of 3.1 (\SI{1.55}{keV}) was measured on a membrane consisting of Al\textsubscript{2}O\textsubscript{3}/TiN/Al\textsubscript{2}O\textsubscript{3} with layer thicknesses of 5/2.5/\SI{5}{\nm}. 

In pursuit of a better performing tynode, ALD magnesium oxide is being considered as membrane material. The choice for MgO stems from the reported reflection electron yield (\gls{REY}) of crystalline MgO, which has a REY of 24.3 (\SI{1.3}{\keV}) \cite{Whetten1957}. It performs better in comparison with Al\textsubscript{2}O\textsubscript{3} (sapphire) and Al\textsubscript{2}O\textsubscript{3} (lucalox), which have a maximum REY of 6.4 (\SI{0.75}{\keV}) and 19.0 (\SI{1.3}{\keV}), respectively \cite{Dawson1966a}. More recently, we have reported the REY of ALD MgO and ALD Al\textsubscript{2}O\textsubscript{3}, which was deposited as thin films on bulk silicon substrates \cite{VanderGraaf2017}. An as-deposited ALD Al\textsubscript{2}O\textsubscript{3} film (\SI{12.5}{\nm}) has a maximum REY of 3.6 (\SI{0.4}{\keV}), whereas an as-deposited ALD MgO film (\SI{15}{\nm}) has a maximum REY of 4.1 (\SI{0.5}{\keV}). We have also shown that thermal and chemical treatment of the thin films can improve the REY \cite{Prodanovic2018b}. For instance, the REY of ALD MgO film (\SI{15}{\nm}) on bulk silicon improved from 4.1 (\SI{0.5}{\keV}) to 5.4 (\SI{0.65}{\keV}) by annealing the film at \SI{700}{\celsius}. Exposure to both high temperatures and chemicals is often part of MEMS fabrication. Fortunately, these treatments seem to be beneficial for our purpose. Additional thermal and/or chemical treatments with the sole purpose to improve the TEY can be considered in the fabrication process as well. 

In ref \cite{Prodanovic2017}, we have implemented ALD MgO as membrane material in a tynode and reported preliminary TEY results. The maximum TEYs of the TiN/MgO films were 2.9 (\SI{1.35}{\keV}), 2.4 (\SI{2.35}{\keV}) and 2.5 (\SI{5.05}{\keV}) for membranes with a thickness of 5, 15 and \SI{25}{\nm}, respectively. The thickness of the TiN layer is 1.8 - \SI{2}{\nm}. However, the membranes seemed to be affected by charge-up effects during the measurement, which resulted in non-smooth yield curves for the membranes with a thickness of 5 and \SI{15}{\nm}. Charge-up effects often diminish the electron yield. The TEY curve of the membrane with a thickness of \SI{25}{\nm} is smooth. The TEY curve is similar to the curve reported by Arntz \& van Vliet \cite{Arntz1962}. They reported a maximum TEY of 2.6 (\SI{3}{\keV}) from a \SI{47.5}{\nm} self-supported MgO film. 

In this work we present a new measurement method that we will use to reexamine the ALD MgO tynodes of ref \cite{Prodanovic2017}. The new method requires a lower electron dose, which can prevent charge-up effects that was observed before. It utilizes the imaging capability of a scanning electron microscope (\gls{SEM}). During image acquisition, the transmission current is measured simultaneously, which is used to construct a TEY map. Using the matching SEM image, the TEYs of individual membranes can be extracted. The TEY of ALD MgO tynodes with membrane thickness of 5 and \SI{15}{\nm} will be re-evaluated using this new method. Also, the collector setup is modified: the semi-spherical grid and collector are replaced by a planar collector. The planar collector is used to investigate the effect of a strong electric field on the TEY of tynodes. A strong electric field can increase the transmission secondary electron yield, as reported by Qin et al. \cite{Qin2011}. Increased secondary electron emission due to electric fields just below the field emission threshold was observed. If a similar enhancement is present in a tynode stack, then the TEY requirement of 4 or higher can be lowered. 

We will present the fabrication process of ALD MgO tynodes in section 2 and the new method to reexamine them in section 3. In section 4, we will first discuss the reliability of the fabrication process and how improvements can be made. Second, we will make a comparison between the method that was used in ref \cite{Prodanovic2017} and the method presented in this work. Third, we will discuss the variance in TEY of individual membranes in an array of a tynode. Lastly, we will discuss the effect of an extracting field on the TEY. In section 5, we will give an outlook of the development of TiPC.


\section{Fabrication}
\label{sec:fab}

\begin{figure}
\centering
\includegraphics[width=0.8\textwidth,origin=c,angle=0]{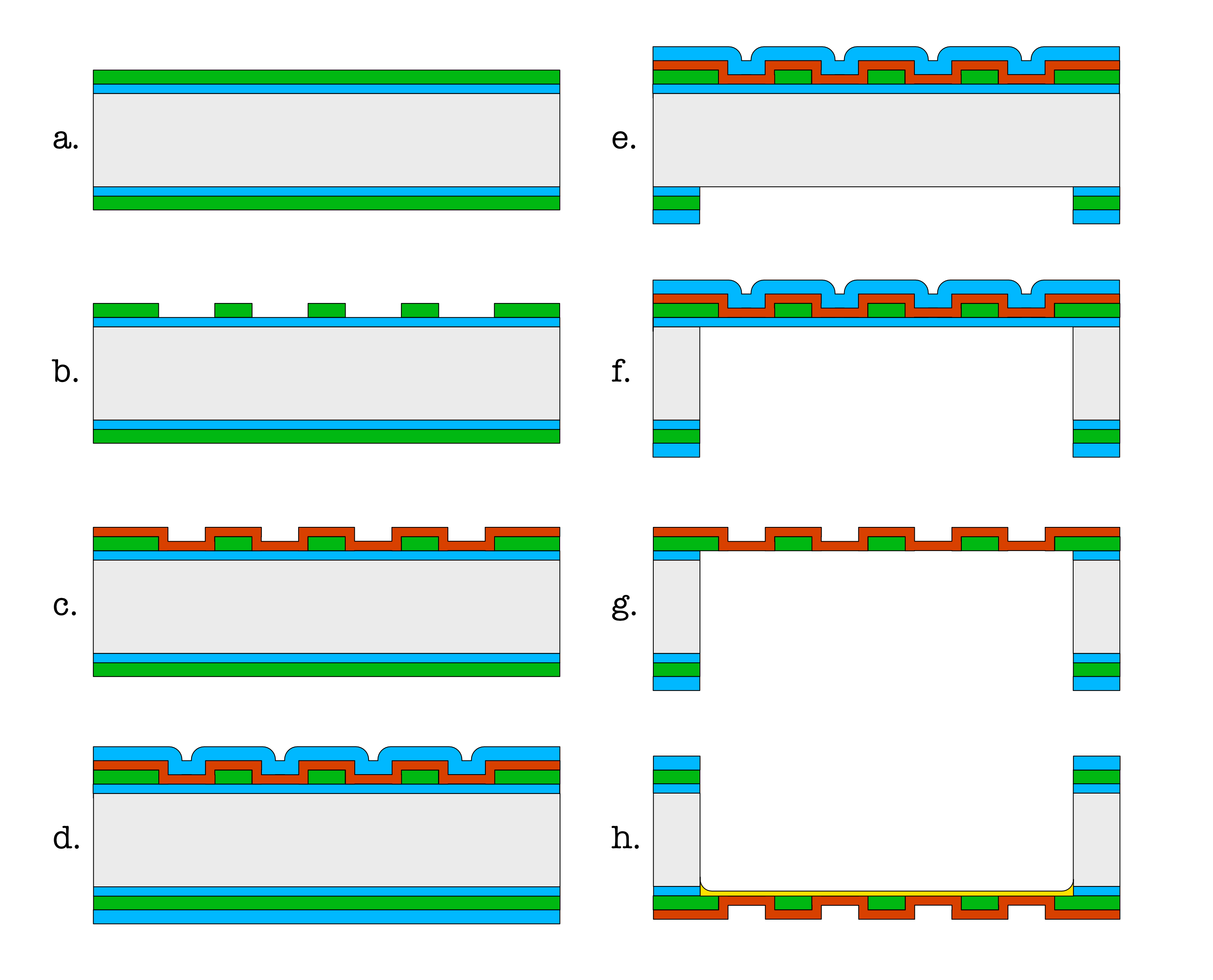}
\caption{\label{fig:3} The fabrication process of a tynode: (a.) Thermal oxidation (\SI{500}{\nm}) and LPCVD of SiN (\SI{500}{\nm}) (b.) Lithography and plasma etch (c.) ALD of MgO (5 or \SI{15}{\nm}) (d.) Plasma-enhanced chemical vapor deposition (PECVD) of silicon dioxide on the front (\SI{1}{\um}) and back (\SI{3}{\um}) (e.) Backside lithography and plasma etch (f.) DRIE etching (g.) HF vapor etching (h.) TiN sputtering}
\end{figure}

The fabrication process of a tynode can be divided in three parts. First, a support mesh of silicon nitride was formed on a silicon wafer (Figure \ref{fig:3}a,b). Second, the ultra-thin film and a protective sacrificial layer of silicon dioxide were deposited on the mesh (Figure \ref{fig:3}c,d). Third, the support mesh and ultra-thin membranes are released (Figure \ref{fig:3}e-h). The fabrication process and the properties of the ALD MgO layer, such as its optical properties, elemental composition and surface morphology, can be found in ref \cite{Prodanovic2017, Prodanovic2019}. 

As substrate, a 4-inch p-type (5 - \SI{10}{\ohm}) silicon wafer with a thickness of ${525}\pm\SI{15}{\um}$ was used. After a standard cleaning procedure, a thermal oxide layer (\SI{500}{\nm}) was grown in a wet thermal environment at \SI{1100}{\celsius}. This layer will act as a stopping and sacrificial layer in the process. On top, a LPCVD silicon nitride layer (\SI{500}{\nm}) was deposited to form the support mesh (Figure \ref{fig:3}a-b). A grid pattern with circular openings with a diameter of \SI{30}{\um} was transferred by photolithography. The pitch between the openings is \SI{55}{\um}, which matches the pixel pitch of a TimePix chip. The SiN in these openings was removed by a plasma etch using hexafluoroethane (C\textsubscript{2}F\textsubscript{6}). 

The wafer was then transferred to a hot wall ALD reactor at \SI{200}{\celsius} (Figure \ref{fig:3}c). As precursors, Mg(Cp)\textsubscript{2} maintained at \SI{80}{\celsius}, and deionized H\textsubscript{2}O at room temperature was used. The gas flow of the N\textsubscript{2} carrier was set to 300 sccm to provide a background pressure of 1 Torr. The precursors and carrier gas were alternately pulsed for a duration of 3, 15, 1 and \SI{15}{\s} for Mg(Cp)\textsubscript{2}, N\textsubscript{2}, H\textsubscript{2}O and N\textsubscript{2}, respectively. The cycle was repeated until the desired thickness was achieved. A PECVD oxide layer (\SI{1}{\um}) was deposited on top of the ALD MgO layer to protect it against subsequent processing steps (Figure \ref{fig:3}d). On the backside, a PECVD oxide layer (\SI{3}{\um}) was deposited that acted as a masking layer for Deep-reactive Ion Etching (\gls{DRIE}). A pattern of large square openings and break lines was transferred by photolithography to the backside of the wafer. First, the PECVD oxide in the openings was removed by a plasma etch (Figure \ref{fig:3}e). Then, the silicon was etched by DRIE (Figure \ref{fig:3}f). The wafer was then separated into individual dies. The individual dies were placed on a carrier wafer for further processing. The support grid and the ultra-thin membranes were released by etching the oxide layers in a hydrogen fluoride (HF) vapor chamber (Figure \ref{fig:3}g) using 4 etching cycles of HF and ethanol with a flow of 190 sccm and 220 sccm, respectively, at 125 Torr. Each cycle had a duration of \SI{10}{\min} and the chamber was purged after each cycle with N\textsubscript{2}. The final step was the deposition of a TiN layer ($\SI{2}{\nm}$) into the opening on the backside onto the silicon support grid and the ultra-thin membranes, which provides lateral conductivity. Silver paint was applied to the silicon substrate as electrical contact points to the sample holder.


\section{Experimental setup}
\label{sec:setup}

\begin{figure}
  \centering
	\subfloat[\label{fig:4a}]{\includegraphics[height=8cm]{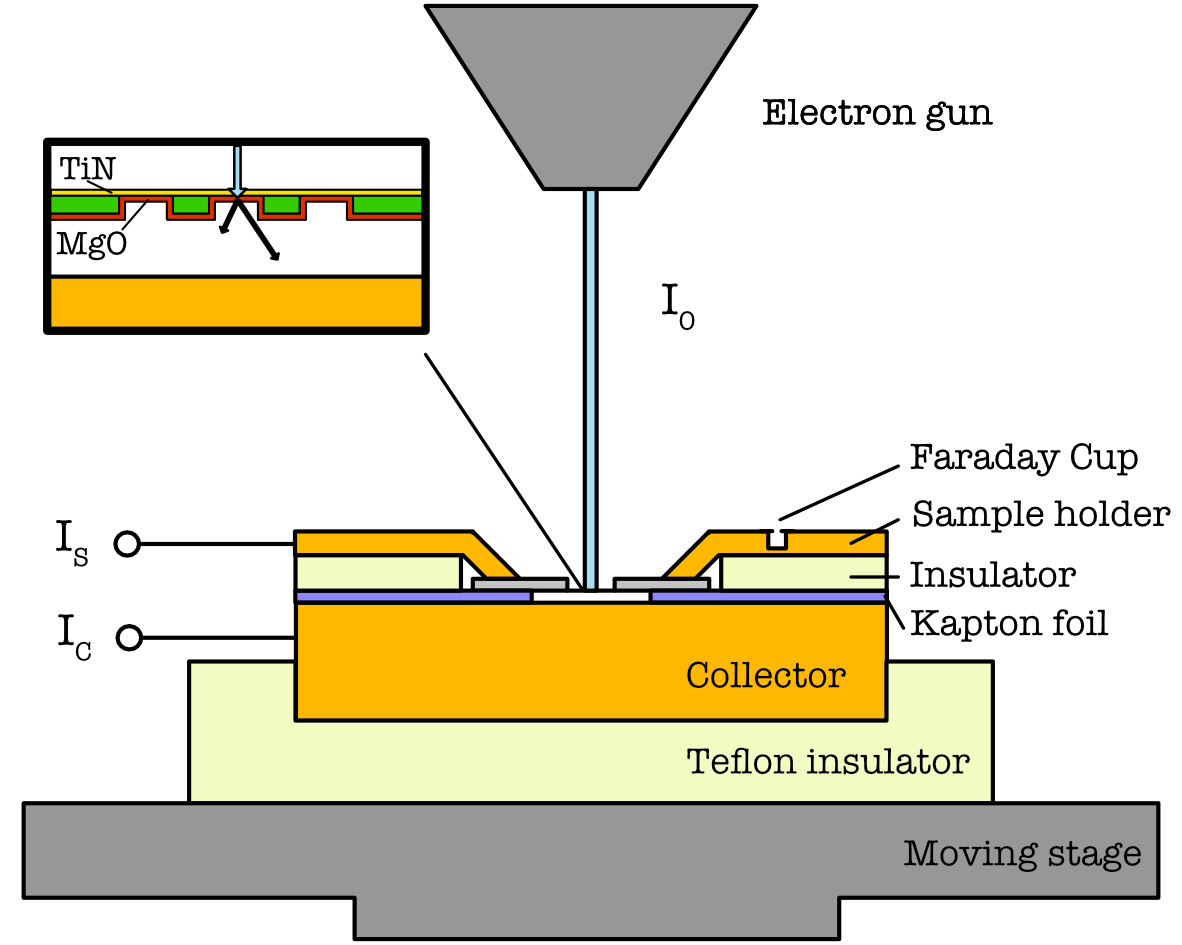}} \
	\subfloat[\label{fig:4b}]{\includegraphics[height=8cm]{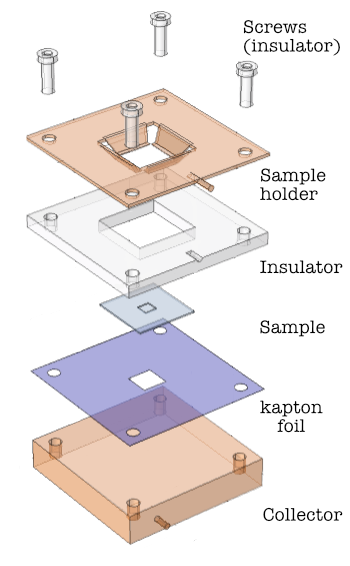}}
     \caption{Experimental setup. \protect\subref{fig:4a} Schematic drawing of the collector system. \protect\subref{fig:4b} Exploded view of the collector assembly} \label{fig:4}
\end{figure}

The transmission electron yield is determined by using a collector-based method within a scanning electron microscope (\gls{SEM}). A collector assembly is mounted on the moving stage of a Thermo Fischer NovaNanolab 650 Dual Beam SEM using a teflon holder (Figure \ref{fig:4a}). The collector assembly has two electrodes, the collector and the sample holder, which are connected to two separate Keithley 2450 sourcemeters via a feedthrough into the SEM chamber. They are electrically insulated from each other using a teflon spacer and teflon screws (Figure \ref{fig:4b}). The sourcemeters can apply a bias voltage to each electrode, ranging from \SI{-200}{\V} to +\SI{200}{\V}, while simultaneously performing a current measurement. The sample is placed on a thin sheet of Kapton (\SI{50}{\um}) with a square opening of \SI{1}{\cm^2} in the center. It is electrically insulated from the collector, but is in contact with the sample holder via folded pressing pins. The distance between the exit surface of the tynode and the bottom electrode can be varied by using Kapton sheets with different thicknesses.  

The sample holder is biased to \SI{-50}{\V} with respect to the collector and the SEM chamber, which are both grounded. In this case, the (secondary) electrons that escape into vacuum are repelled from the sample and sample holder. In this work, we do not distinguish the fast electrons, backscattered and forward-scattered, from the slow secondary electrons. The transmission electrons are collected by the collector, while reflection electrons are absorbed by the chamber. The beam current $I_0$ depends on the primary energy $E_0$ of the beam and is measured with a Faraday cup. Before and after a measurement, the leakage and background currents in the setup are measured, which are subtracted from the collector and the sample current. The actual landing energy of the PEs is lower due to the negative bias on the sample with respect to the SEM and should be corrected by \SI{50}{\eV}. The electron dose of a beam on a single spot can be quite high, which can give rise to charge-up effects and surface contamination. Therefore, a scanning electron beam is preferred to distribute the electrons over a larger surface \cite{Chan2021}. The TEY is determined by measuring the transmission current and is given by
\begin{equation}
\label{eq:1}
\sigma_T(E_0) = \frac{I_C}{I_0}
\end{equation}
where $I_C$ is the collector current. The total electron emission is the sum of the emissions on both sides of the membrane: $\sigma(E_0)=\sigma_R(E_0) +\sigma_T(E_0)$, and is given by  
\begin{equation}
\label{eq:2}
\sigma(E_0) = \frac{I_0-I_S}{I_0}
\end{equation}
where $I_S$ is the sample current. The reflection electron yield (REY) is then given by
\begin{equation}
\label{eq:3}
\sigma_R(E_0) = \frac{I_0-I_S-I_C}{I_0}
\end{equation}

In this work, we present a method that allows us to determine the TEY of multiple membranes in the support grid. During the acquisition of a SEM image, the transmission current is measured as a function of time. From this current the TEY is determined using formula \ref{eq:1} and a TEY map is constructed using the same principles as for the SEM image construction. A SEM image is acquired by scanning an electron beam on the surface of the specimen. A surface area corresponding with one pixel in the image is irradiated by an electron beam for the duration of the dwell time. The line time is the required time to acquire one row of pixels plus the time needed to reposition the beam to the next row. The frame time is the acquisition time of the whole image. For this method, a set of SEM parameters is chosen that is optimized to the fastest acquisition rate of a Keithley 2450 sourcemeter, which is the limiting factor. A larger dwell time can be considered, but that would increase the electron dose. The image, with a resolution of 512 x 442, is acquired using a dwell-, line and frame time of \SI{1}{\ms}, \SI{560}{\ms} and \SI{4.2}{\min}, respectively. This timing information is used to divide the measured TEYs in intervals of \SI{560}{\ms} to obtain the rows of pixels that will be used to construct the TEY map. The sourcemeter has an acquisition rate of \SI{333}{\s^{-1}} or a sample time of \SI{3.3}{\ms}, so each row will correspond to only 168 data points. The missing pixels in the TEY map are interpolated to obtain a map with a resolution of 512 x 442. The individual membranes are identifiable on this map and the TEY of each membrane can be extracted. Likewise, the REY map can be obtained using this method. 

The collector assembly with the planar collector is designed to investigate the effect of the electric field strength on the TEY. In the design of TiPC, the tynodes are separated by insulating spacers. The distance between two tynode membranes will be approximately \SI{600}{\um}, which is the substrate thickness plus the height of the spacer. With a bias voltage of \SI{1000}{\V}, the electric field will be \SI{1.67e6}{\V/\m}. In this measurement, the electric field will be in the same order of magnitude as the intended operating conditions of TiPC. The distance between the emission surface and the collector depends on the thickness of the Kapton foil, which is \SI{50}{\um}. In a standard measurement, the bias voltage between the sample and collector is \SI{50}{\V}, which gives an electric field of \SI{1e6}{\V/\m}.


\section{Results and discussions}
\label{sec:results}

\subsection{Tynode fabrication}
\label{ssec:tynode}

\begin{figure}
\centering
\includegraphics[width=0.8\textwidth,origin=c,angle=0]{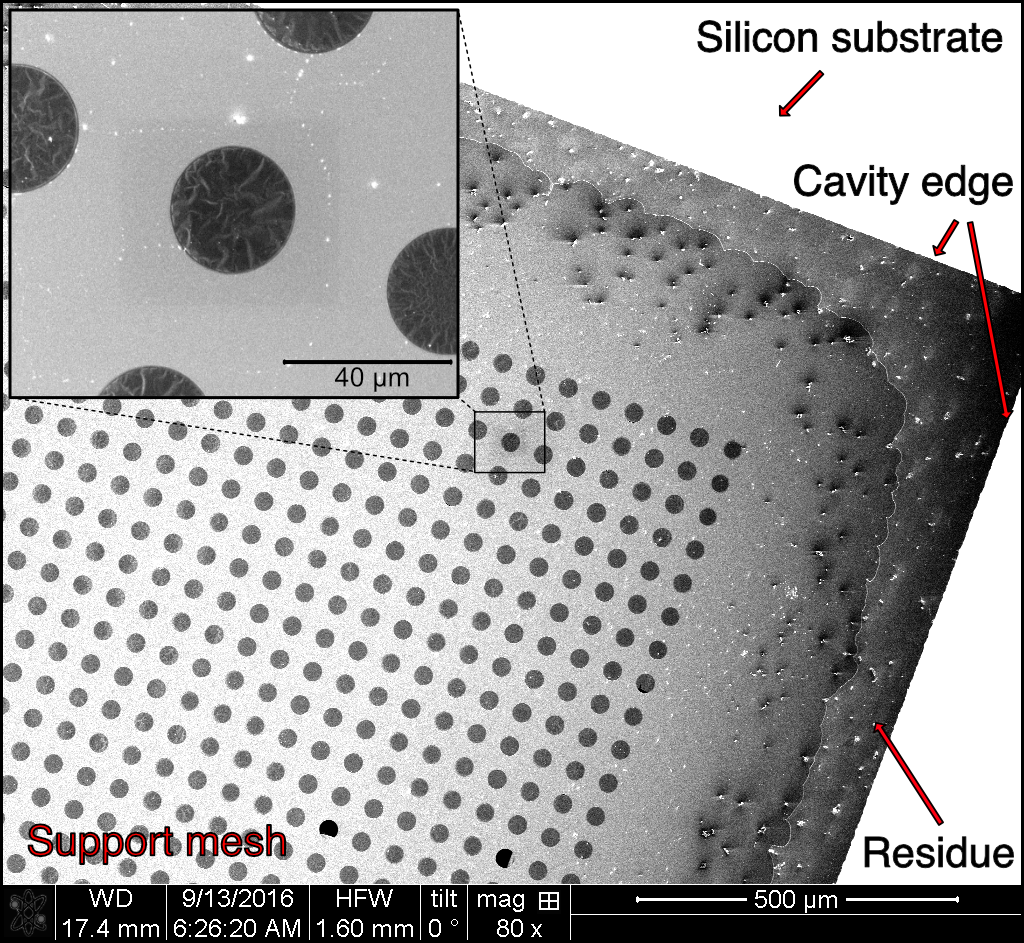}
\caption{A SEM image of a tynode acquired with 2-keV-electrons using a magnification of 80x and 1200x (Overlay image). The tynode membranes have a thickness of \SI{5}{\nm} and a diameter of \SI{30}{\um}. The SEM image depicts the backside of a tynode in which a large square cavity is etched into the silicon substrate. The cavity edge is clearly visible. The support grid in which the ultra-thin membranes are suspended is fully released. Near the side of the cavity, there are some residues, which are the remains of the sacrificial silicon dioxide layer. In the array, the ultra-thin membranes are also fully released. They appear to be translucent in comparison with the support grid, which is expected since 2-keV-electrons have sufficient energy to penetrate the membranes. As a result, less back-scattered and reflection secondary electrons are reemitted from the membranes.} \label{fig:5} 
\end{figure}

\begin{figure}[h]
  \centering
	\subfloat[\label{fig:6a}]{\includegraphics[height=4.5cm]{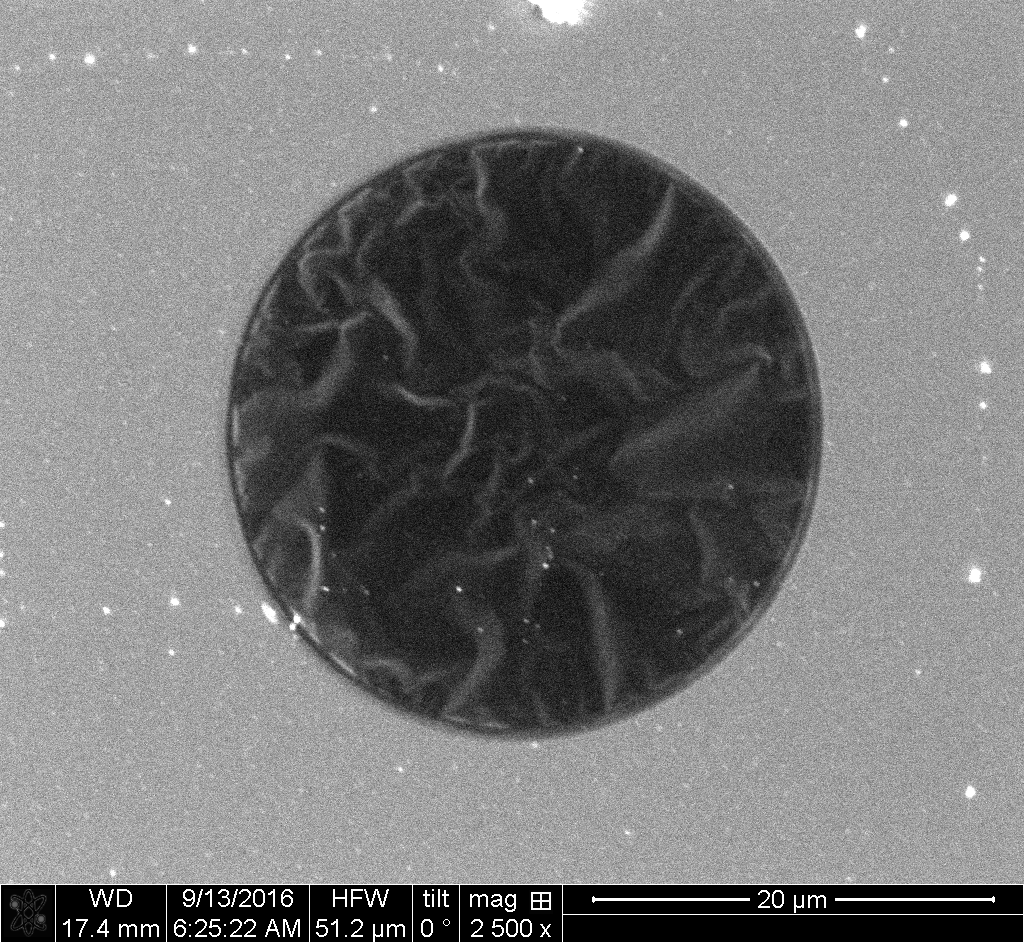}} \
	\subfloat[\label{fig:6b}]{\includegraphics[height=4.5cm]{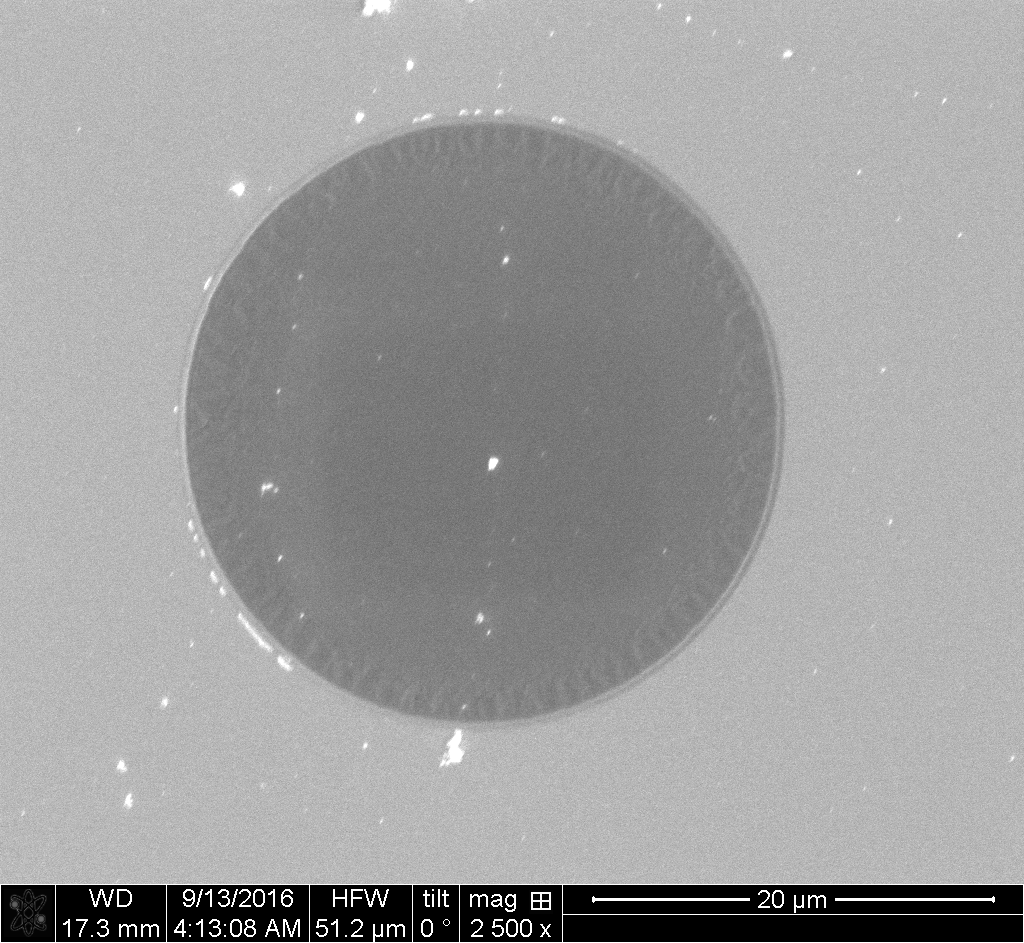}} \
	\subfloat[\label{fig:6c}]{\includegraphics[height=4.5cm]{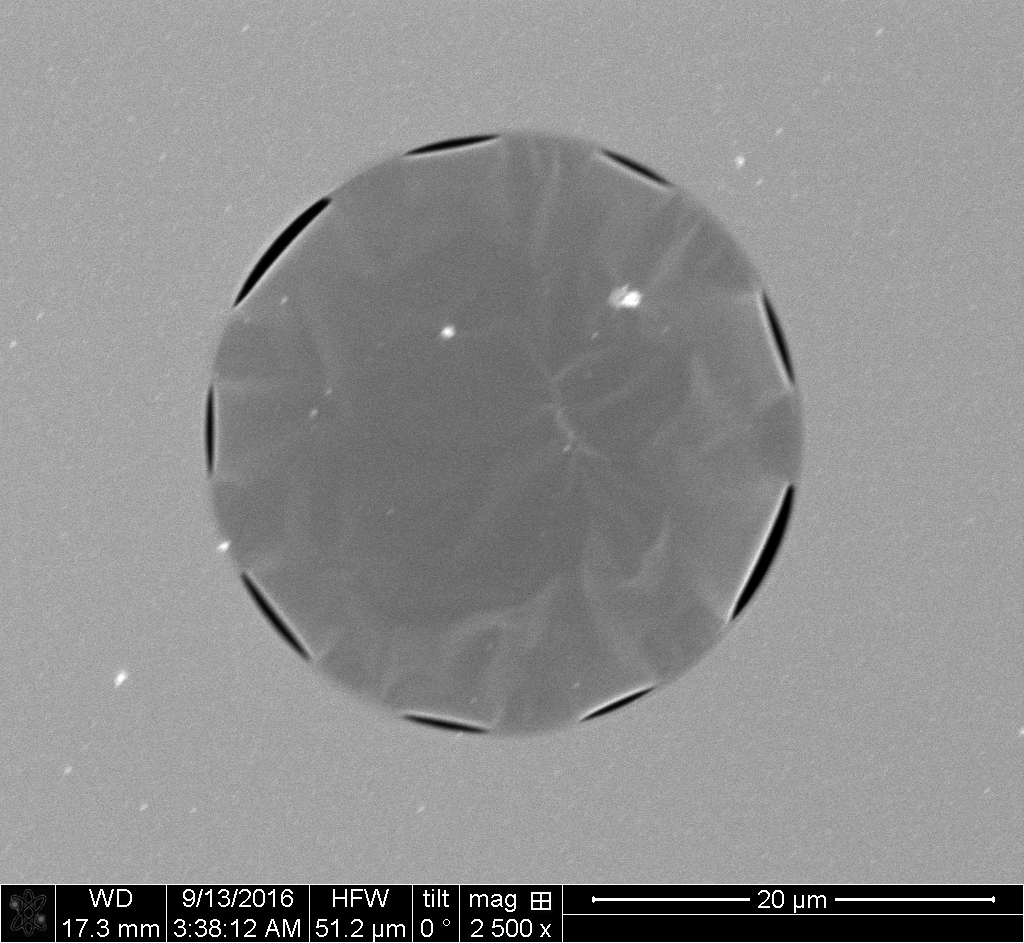}} \	
     \caption{SEM images of various membranes on a tynode acquired with 2-keV-electrons using a magnification of 2500x. The membranes across the array differ in shape, as \protect\subref{fig:6a} some are buckled while \protect\subref{fig:6b} others are flat and \protect\subref{fig:6c} some shows tears near the edge. The difference in contrast is due to the SEM setting used to acquire the image and not necessarily an intrinsic difference between the membranes. Other variations are due to minute differences in etch time and conditions to which the membranes are subjected depending on their position in the array.} \label{fig:6}
\end{figure}

An image of a tynode is acquired with a SEM using an electron beam energy of $E_0=\SI{2000}{\eV}$ (Figure \ref{fig:5}). The contrast in the image is due to the different materials that are present on the tynode. Also, in the case of ultra-thin membranes, the thickness plays a role in the contrast, since the PEs have sufficient energy to penetrate the membranes. As a result, a thinner membrane backscatters less PEs. The PEs generate less reflection SEs, which results in a darker appearance on the image. In figure \ref{fig:5}, the membranes consist of a layer of MgO with a thickness of \SI{5}{\nm} and a layer of TiN with a thickness of \SI{2}{\nm}. The TiN layer was sputtered as a post-process on an uneven surface, which could lead to a less uniform deposition in comparison with a deposition on a flat wafer surface. 

The ultra-thin membranes differ slightly in shape across the array (Figure \ref{fig:6}). There are some that are buckled, some that have tears near their edges and others that are flat. Buckling and tearing are due to residual stress in the membrane, which can vary across the array. There are minute differences in etch and deposition rates that depend on the surface topography. For instance, the etch rate of DRIE is higher in the center of a square cavity. The silicon is removed quicker and more landing material (silicon dioxide) is removed from the center of the cavity. As a result, the MgO membranes in the center are exposed to HF vapor for a longer duration in comparison to the membranes near the edges. Also, the deposition rate for sputtering is higher in the center of a cavity. The composition of the membranes may vary as well as their residual stress. 

In a different study, the effects of HF vapor etching on ALD MgO films have been investigated using X-ray photoelectron spectroscopy (\gls{XPS}) \cite{Prodanovic2018b}. The XPS data showed that fluorine was present on the surface as well as in the bulk of thin MgO films after exposure to HF vapor. The samples were prepared by depositing thin ALD MgO films on silicon substrates. Some of the films were encapsulated with a protective oxide layer to mimic the fabrication processes involved in the present work. The samples were subsequently exposed to HF vapor for a short duration or until the oxide layers were removed. An in-depth XPS analysis of atomic content was performed by argon-ion-sputtering of the films. The results indicate that some of the oxygen atoms were exchanged for fluorine and that MgF\textsubscript{2}/MgO compounds had formed. Moreover, the electron emission properties of the films changed. In all cases, the reflection SE yields improved. It is likely that fluorine is present in the MgO membranes of tynodes. 
 

\subsection{Surface scan method}
\label{ssec:surface}

\begin{figure}
  \centering
\includegraphics[width=1\linewidth,origin=c,angle=0]{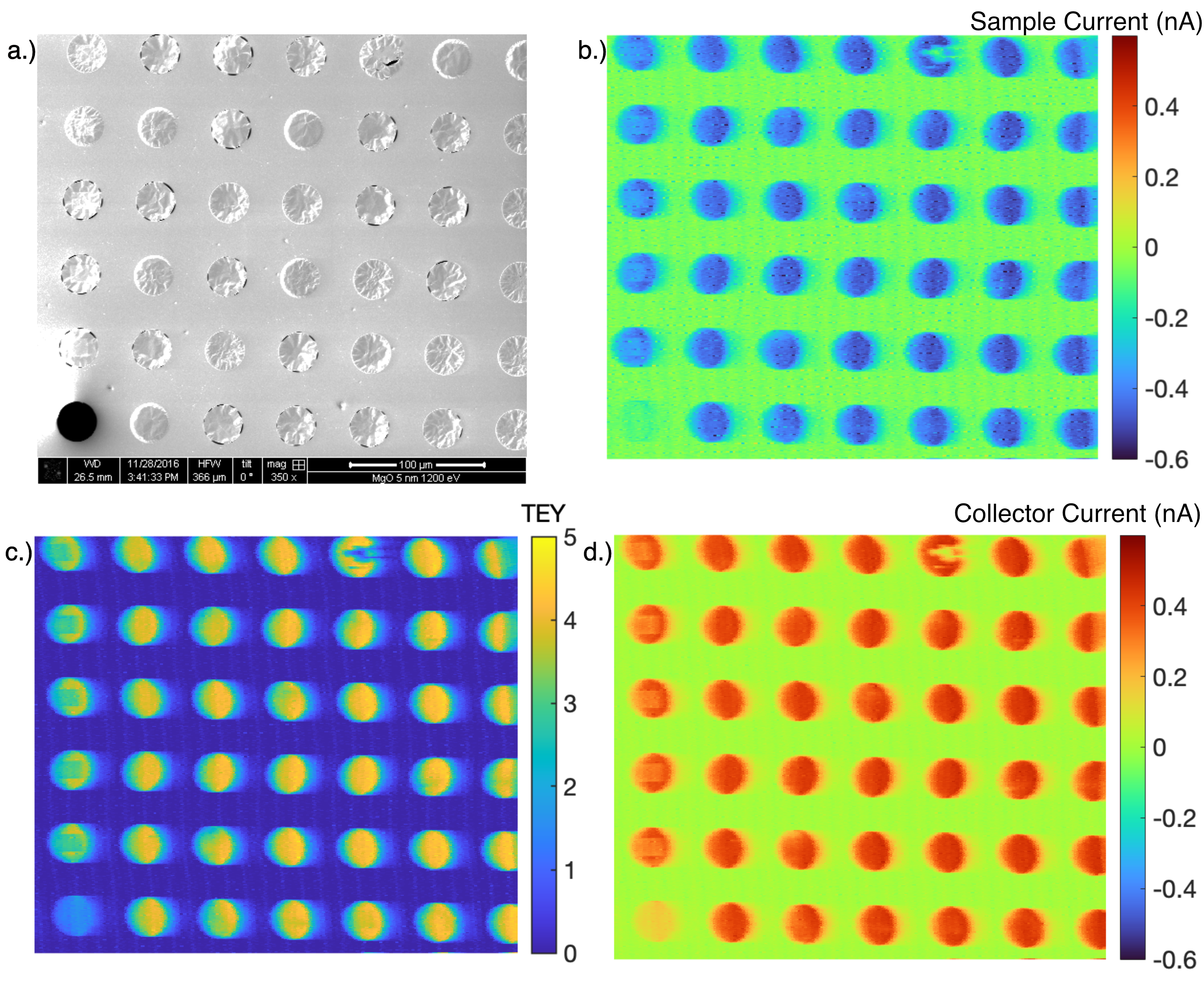}
\caption{\label{fig:7} A TEY map (re)construction using the measured currents as a function of time. (a) A SEM image with a resolution of 512 x 442 pixels is acquired with a dwell time of \SI{1}{\ms} using 1.2-keV-electrons. (b), (d) The sample current and collector current are used to construct maps from which the emission currents of individual membranes in the array can be extracted to determine their REYs and TEYs. (c) A TEY map that is determined by using formula \ref{eq:1}.}
\end{figure}

\begin{figure}
  \centering
	\subfloat[\label{fig:8a}]{\includegraphics[height=9cm]{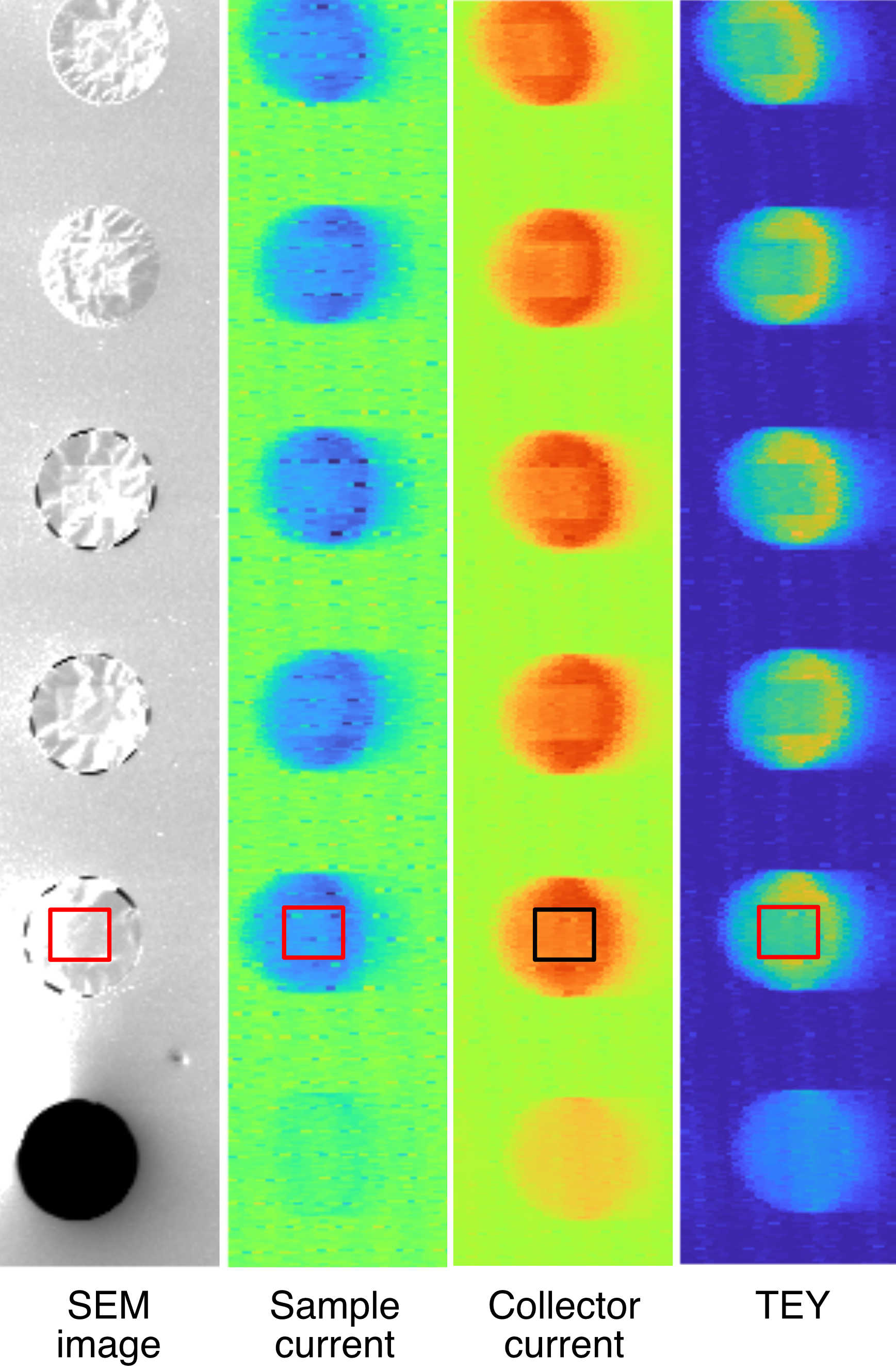}} \
	\subfloat[\label{fig:8b}]{\includegraphics[height=9cm]{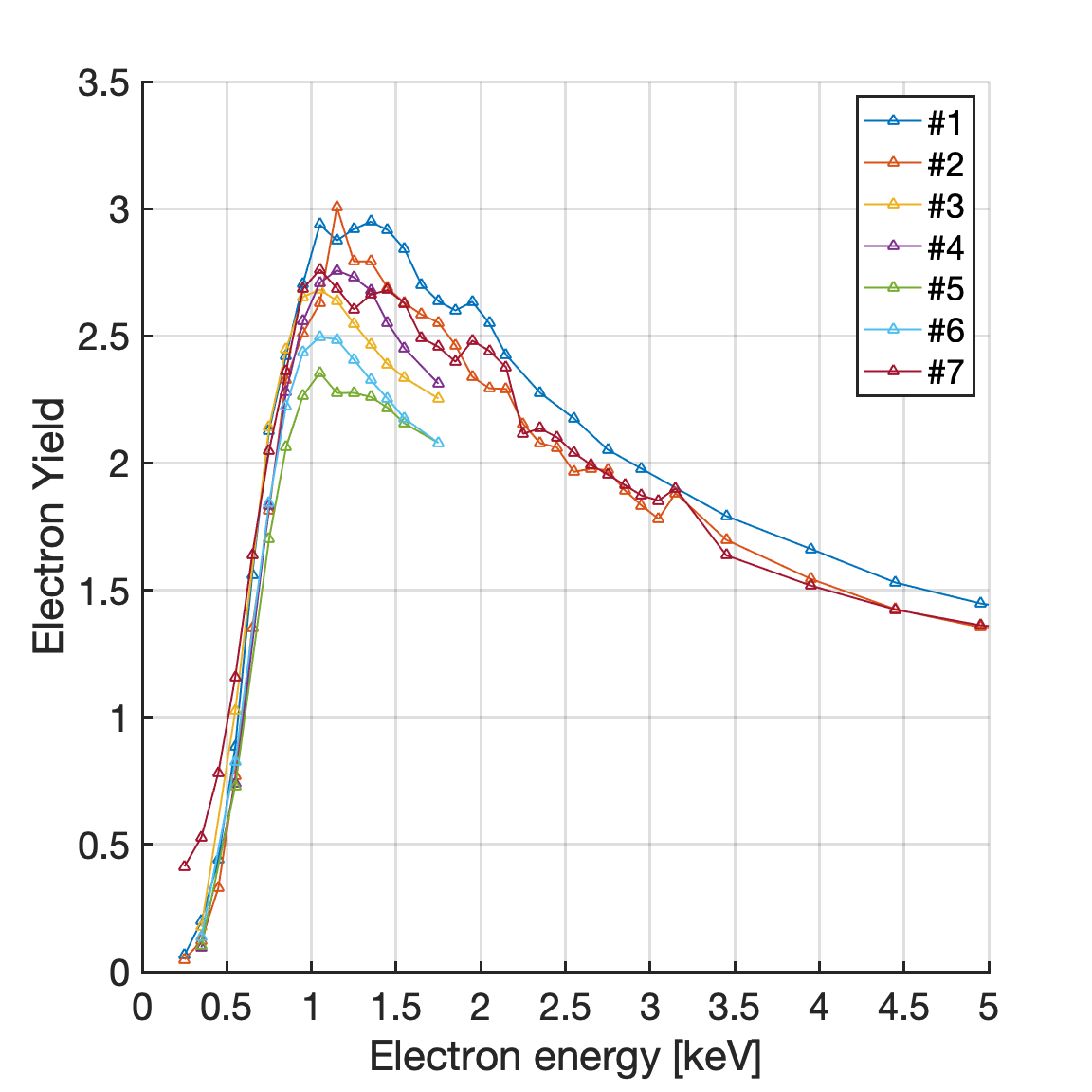}} \
     \caption{The effect of prolonged electron irradiation on the TEY. \protect\subref{fig:8a} A close-up of the first column of the array in figure \ref{fig:7}. The first column was subjected to a series of measurements using a different measurement method, which left imprints on the membranes. Individual membranes were irradiated with a scanning electron beam. The irradiated surfaces coincide with the rectangular discolorations in the centers of the membranes. The time between this series of measurement and the SEM image is 16 days, so the imprints on the membranes seem to be permanent as well as their reductions in TEYs. A more extensive description of the method can be found in ref \cite{Chan2021}. \protect\subref{fig:8b} The TEY curves of 7 individual membranes that were obtained using the method described in ref \cite{Chan2021}. The TEY curves are non-smooth, which can be due to charge accumulation or electron beam induced contamination. Both effects can change the electron emission properties of the membrane during the measurement, which result in artefacts in the TEY curves.} \label{fig:8}
\end{figure}

The surface scan method has two advantages in comparison with methods that use an electron gun with a static beam. First, the SEM provides an image of the surface that allows us to determine the electron emission of multiple membranes on a tynode simultaneously. Second, the electron dose that is subjected to the surface is lower in comparison with a direct beam, which will minimize charge-up effects and/or build-up of contamination. 

In figure \ref{fig:7}, the surface scan method was applied to a section of a tynode. A SEM image with a resolution of 512 x 442 pixels was acquired using a magnification of 350x, a dwell time of \SI{1}{\ms} and a beam energy of \SI{1.2}{\keV} (Figure \ref{fig:7}a). Simultaneously, the sample and collector currents were measured as a function of time. By splitting the currents in segments that correspond with the rows of the SEM image and placing them in order, a map that matches the SEM image was constructed (Figure \ref{fig:7}b-c). From this map, the measured currents were extracted for individual membranes and their REYs and TEYs were determined using formula \ref{eq:1} and \ref{eq:3}. A TEY map was drawn by dividing the collector current by the beam current, which was \SI{0.101}{\nA} for 1.2-keV-electrons (Figure \ref{fig:7}d).

As seen in figure \ref{fig:7}, one of the membranes broke off during the fabrication process. The opening contrasts sharply with its surroundings and was used as a reference point on the grid. Surprisingly, a sample current was measured on that position during a surface scan. The PEs should pass through the opening and land on the collector directly. The collector current would be equal to the beam current and a TEY of 1 was expected. However, the measured collector current was larger than the beam current. This discrepancy can be attributed to backscatter electrons from the collector surface. The backscattered electrons generate reflection SEs on the backside of the tynode/membranes, which will subsequently be absorbed by the collector. In this case, the measured sample current is due to this 'secondary' process. When a membrane is present, this 'secondary' process will also occur for PEs with sufficient energy to penetrate the membrane and backscatter from the collector. 

As mentioned, the electron dose that is applied to the surface is lower in comparison with static beam setups. It is also lower in comparison with the previous method that we have used in ref \cite{Prodanovic2017}. In figure \ref{fig:8a}, a consequence of a high electron dose can be seen. The first column was subjected to a series of measurements in which the electron beam targeted single membranes. The membranes have discolorations at their centers. A rectangle is vaguely visible on the SEM image, the sample and collector current maps. The rectangle corresponds  with the surface area that was imaged by the SEM using a magnification of 8000x. The electron dose to obtain one yield curve (see figure \ref{fig:8b}) is approximately \SI{9e-3}{\coulomb/\um^{2}}. In comparison, using the method presented in this work, the electron dose to obtain a TEY curve (see figure \ref{fig:9b}) is \SI{1.96e-4}{\coulomb/\um^{2}}, which is 46 times lower. Apparently, prolonged exposure to electron bombardment reduces the electron emission of these membranes. There are two mechanisms that can lead to a reduction in electron yield. First, when an insulator is irradiated by an electron beam, a positive charge can accumulate if there is insufficient conductivity to replenish the emitted electrons. As charge builds up, the electric field with increasing strength will retract more SEs until it reaches an equilibrium where the net yield is one. The second mechanism is the buildup of electron-beam-induced contaminations. Hydrocarbons on the surface of a sample that are being irradiated by an electron beam will start to form a contamination layer. The secondary electron emission properties of this contaminant differ from MgO. Of these two mechanisms, one is reversible and one is permanent. In the first case, there are mechanisms to discharge the trapped charge, whereas in the second case, once the contaminant is formed, it cannot be removed. Therefore, the electron dose that is used to inspect the samples should be kept at a minimum.

\subsection{Transmission electron yield}
\label{ssec:TEYcurve}

\begin{figure}
  \centering
	\subfloat[\label{fig:9a}]{\includegraphics[height=7.2cm]{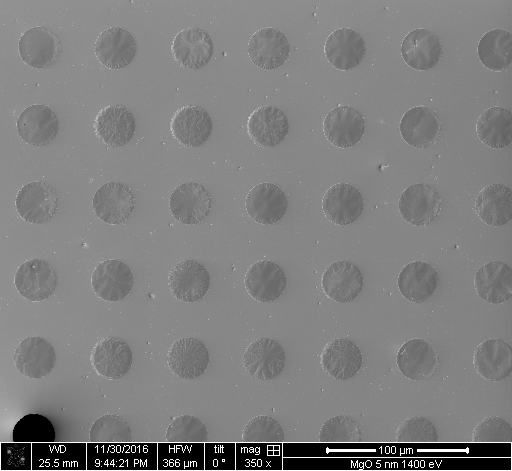}} \
	\subfloat[$d=\SI{5}{\nm}$; $\diameter=\SI{30}{\um}$; $n=20$ \label{fig:9b}]{\includegraphics[height=7.2cm]{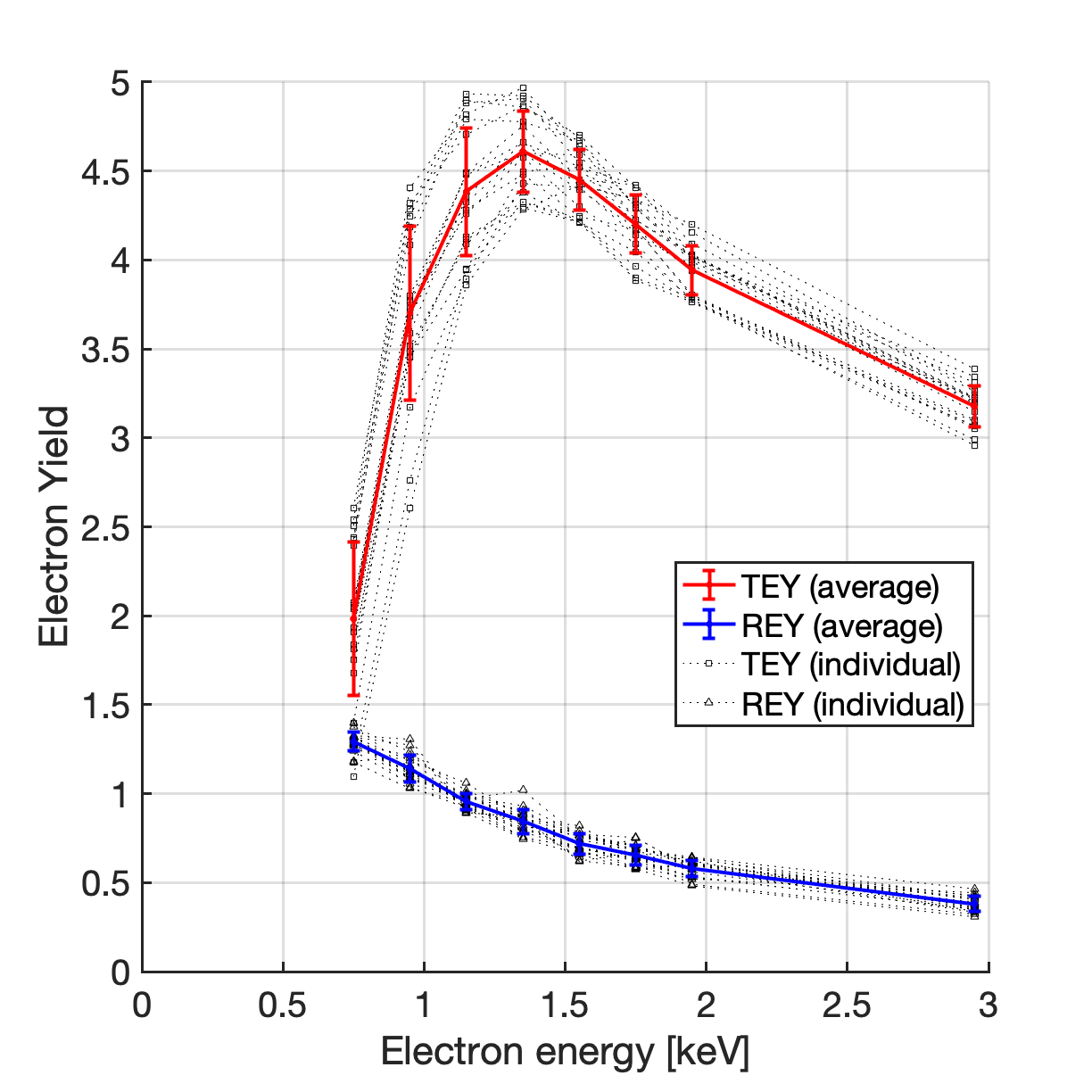}} \
     \caption{The averaged TEY curve of multiple membranes. \protect\subref{fig:9a} A SEM image of membranes with a thickness of \SI{5}{\nm} and a diameter of \SI{30}{\um} acquired with 1.4-keV-electrons. \protect\subref{fig:9b} A set of $n=20$ TEY curves is determined for individual membranes in the SEM image. The spread in TEY is the largest for low electron beam energy. The onset of the TEY curve depends on the membrane thickness, which can vary across the array due to the fabrication processes. The spread in REY is smaller, since the membrane thickness does not affect reflection secondary electron emission as much.} \label{fig:9}
\end{figure}

\begin{figure}
  \centering
	\subfloat[\label{fig:10a}]{\includegraphics[height=7.2cm]{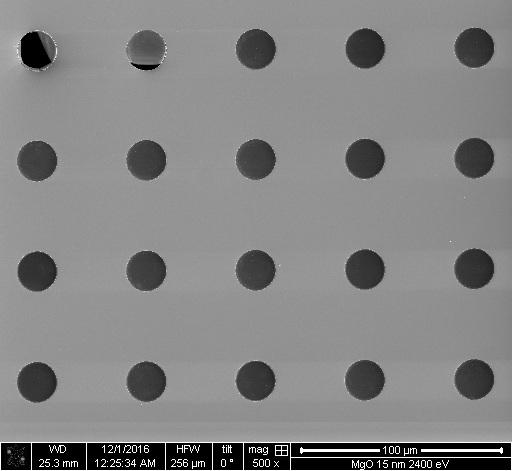}} \
	\subfloat[$d=\SI{15}{\nm}$; $\diameter=\SI{20}{\um}$; $n=18$ \label{fig:10b}]{\includegraphics[height=7.2cm]{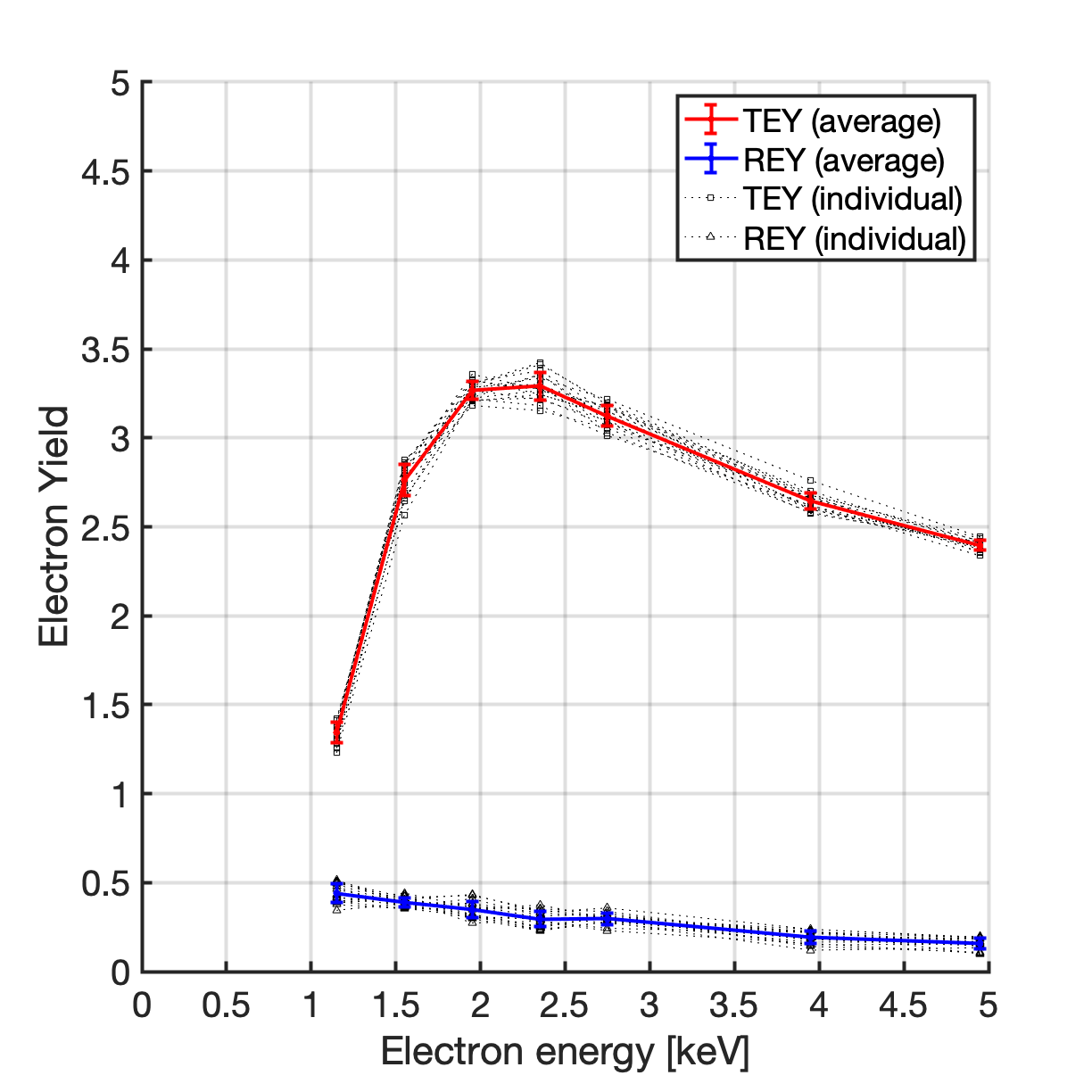}} \
     \caption{The averaged TEY curve of multiple membranes. \protect\subref{fig:10a} A SEM image of membranes with a thickness of \SI{15}{\nm} and a diameter of \SI{20}{\um} acquired with 2.4-keV-electrons. \protect\subref{fig:9b} A set of $n=18$ TEY curves is determined for the individual membranes in the SEM image. The spread in TEY of the \SI{15}{\nm} membranes is smaller compared to the \SI{5}{\nm} membranes. The small variations in the thickness due to the fabrication processes play a relatively smaller role for thicker membranes.} \label{fig:10}
\end{figure}



The transmission electron yields of tynodes with two different thicknesses have been determined using the surface scan method. In figure \ref{fig:9a}, a different section of the tynode with membranes with a thickness $d=\SI{5}{\nm}$ and a diameter $\diameter=\SI{30}{\um}$ is shown. This section has not been irradiated by an electron beam before the measurement. The TEYs of 20 individual membranes have been determined using the surface scan method for electron beam energies of 0.75 - \SI{2.95}{\keV}. Their individual TEY curves and their average (red line) are shown in figure \ref{fig:9b}. An averaged maximum TEY of 4.6 was achieved with a beam energy of \SI{1.35}{\keV}. In figure \ref{fig:10a}, a tynode with membranes with $d=\SI{15}{\nm}$ and $\diameter=\SI{20}{\um}$ is shown. The TEYs of 18 membranes have been determined for primary energies of 1.15 - \SI{4.95}{\keV}. The averaged maximum TEY was found to be 3.3 (\SI{2.35}{\keV}).  


The maximum TEY of the membranes with $d=\SI{5}{\nm}$ is higher than was reported before in ref \cite{Prodanovic2018}, which was 2.9 (\SI{1.35}{\keV}). In figure \ref{fig:8b}, the TEY curves of additional membranes on the same tynode are shown, which were determined using the same method as described in ref \cite{Prodanovic2018}. Each curve is obtained by repeatedly irradiating a single membrane with increasing beam energy. Overall, the maximum TEYs is lower using this method compared to the surface scan method. This can be attributed to the relatively high electron beam intensity, which contaminated the surface as can be seen in figure \ref{fig:8a}. Also, the irregularities in the TEY curves are indications that some charge-up effects were present. Another factor is the presence of a strong electric field near the exit surface using the new setup, which can lower the electron affinity and increase the TEY. We will discuss this in section \ref{ssec:field}.


The variance in the TEY can be attributed to two major factors: the sensitive nature of secondary electron emission, and small variations in film thickness. The first factor has been investigated extensively for reflective secondary electron emission: the experimental conditions and surface termination on the material affect secondary electron emission. As such, the reported REY of many materials can vary a lot. In our case, electron induced contamination can occur when the sample is examined within the SEM. Also, the handling of the sample in ambient conditions might affect the surface condition. The second factor, the small variations in film thickness, is due to the minute differences in etch and deposition rates in the fabrication process due to surface topography. Particularly, the post-process deposition of TiN on the released membranes might result in a less uniform coverage due to the topography of the surface (figure \ref{fig:3}h). 

For the membranes with $d=\SI{5}{\nm}$, there is a large variance in TEY between the membranes, especially for lower primary beam energies (figure \ref{fig:9b}). These energies coincide with the onset of transmission (secondary) electron emission, which occurs when PEs deposit energy near the exit surface of the membrane. A small increase in the thickness would require a slightly higher PE energy to reach the exit the surface. Therefore, small variations in thickness would lead to a large variance in the TEY. For higher PE energy, the majority of the PEs will be transmitted through the membrane. Their energy loss profiles would be similar regardless of the thickness of the membrane. For the membranes with $d=\SI{15}{\nm}$, the variance in TEY is much smaller (figure \ref{fig:10b}). The TEY curves rise more gradually for thicker membranes. A small variation in thickness due to the TiN deposition contributes relatively less to the overall thickness of the membrane. A shift of the onset of the TEY curve will be less apparent in this case. Further, the variance in the REY is small regardless of the membrane thickness. For reflection secondary electron emission, only PE interactions at a shallow depth contribute. The energy loss profile near the surface will be similar regardless of small variations in thickness. 

\subsection{Extraction field-enhanced yield}
\label{ssec:field}

\begin{figure}
  \centering
	\subfloat[$E_0=\SI{1.4}{\keV}$; $d=\SI{5}{\nm}$; $\diameter=\SI{30}{\um}$; $n=25$ \label{fig:11a}]{\includegraphics[height=8.5cm]{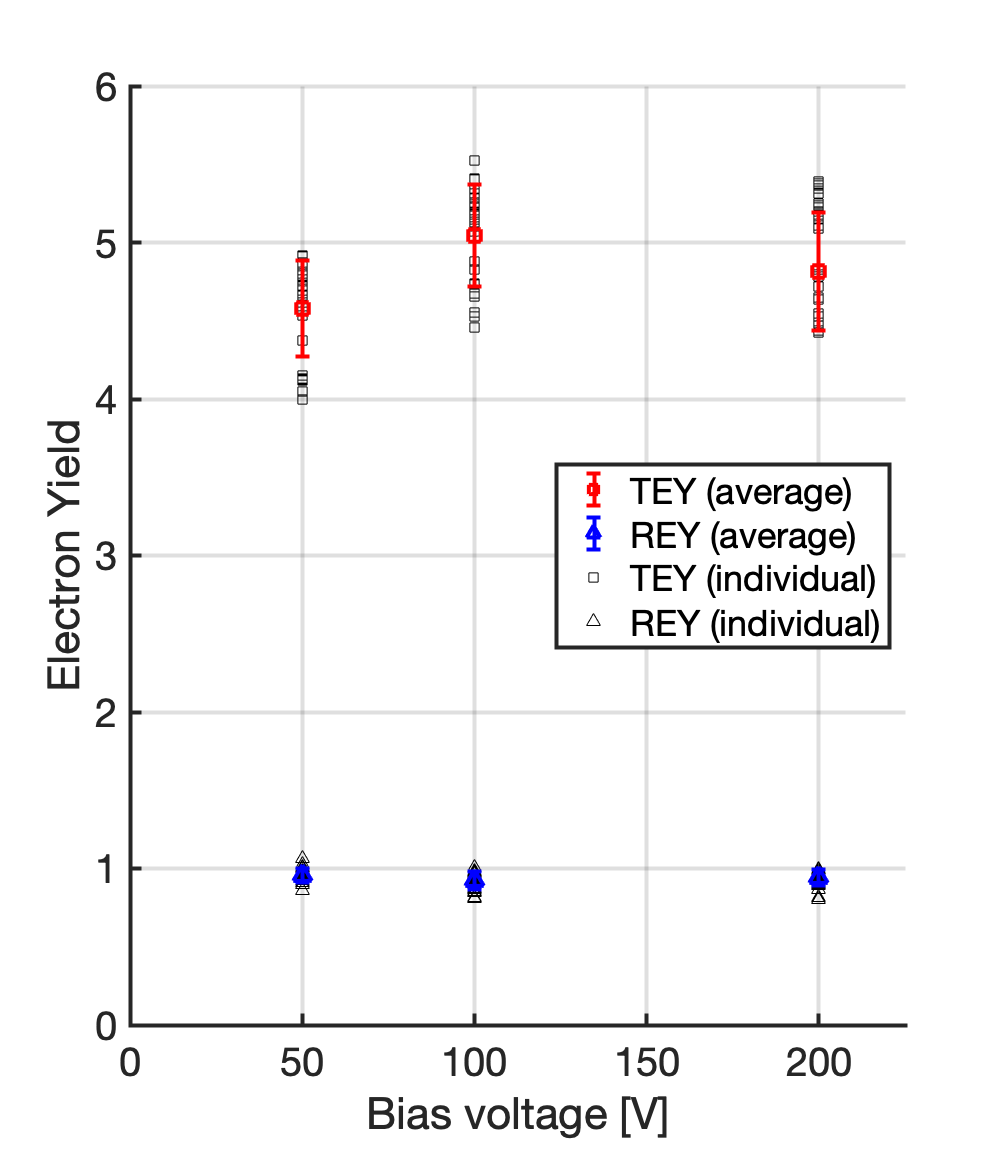}} \
	\subfloat[$E_0=\SI{2.4}{\keV}$; $d=\SI{15}{\nm}$; $\diameter=\SI{20}{\um}$; $n=9$ \label{fig:11b}]{\includegraphics[height=8.5cm]{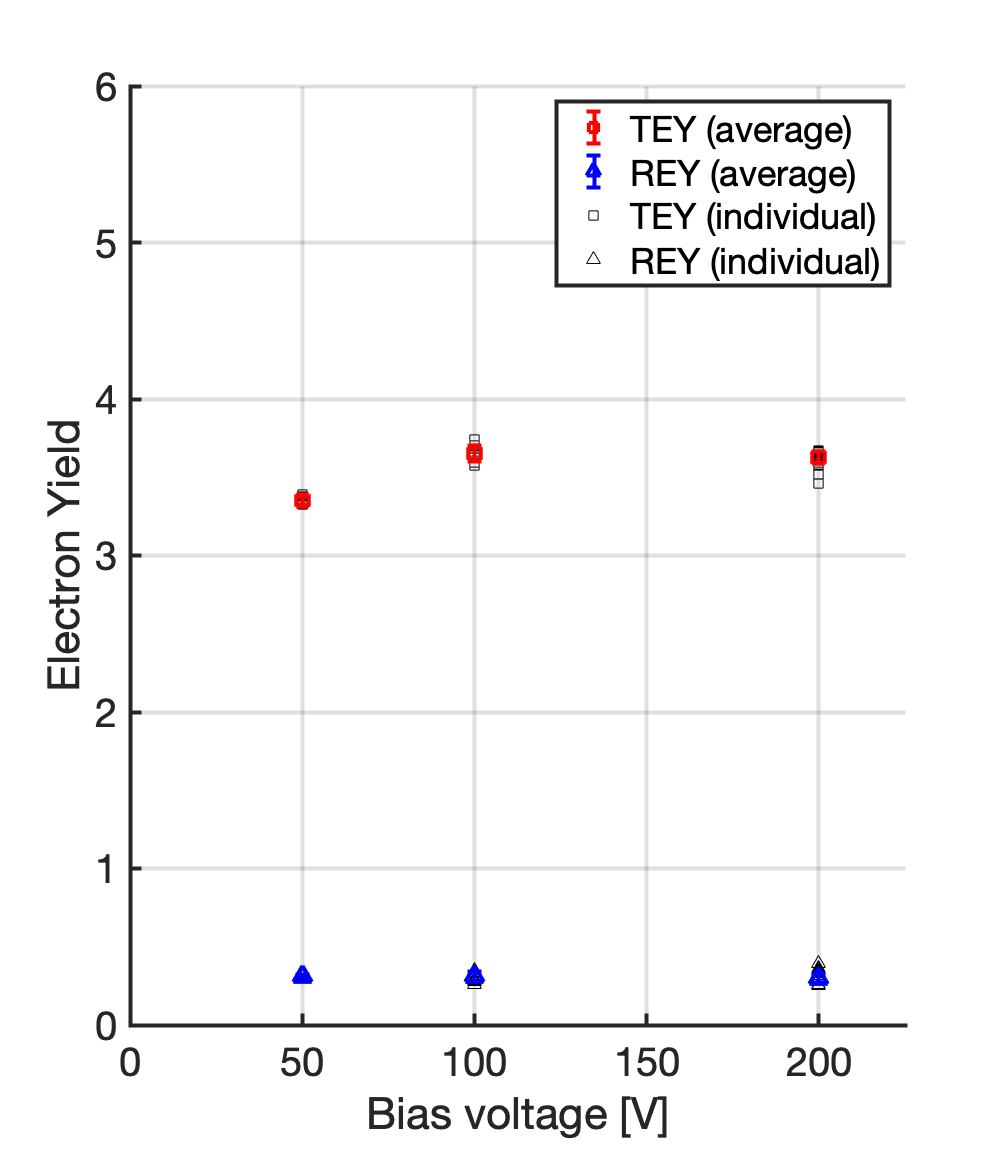}} \
     \caption{The effect of the electric field on the TEY. The electric potential between the sample and collector is increased by setting the collector potential at 0, +50 and +\SI{150}{\V} for three subsequent measurements, while keeping the sample potential at -\SI{50}{\V}. The averaged TEY was determined for two tynodes with $d=\SI{5}{\nm}$ and $d=\SI{15}{\nm}$ using their optimal PE energy $E_0$.} \label{fig:11}
\end{figure}

\begin{figure}
  \centering
\includegraphics[width=0.8\linewidth,origin=c,angle=0]{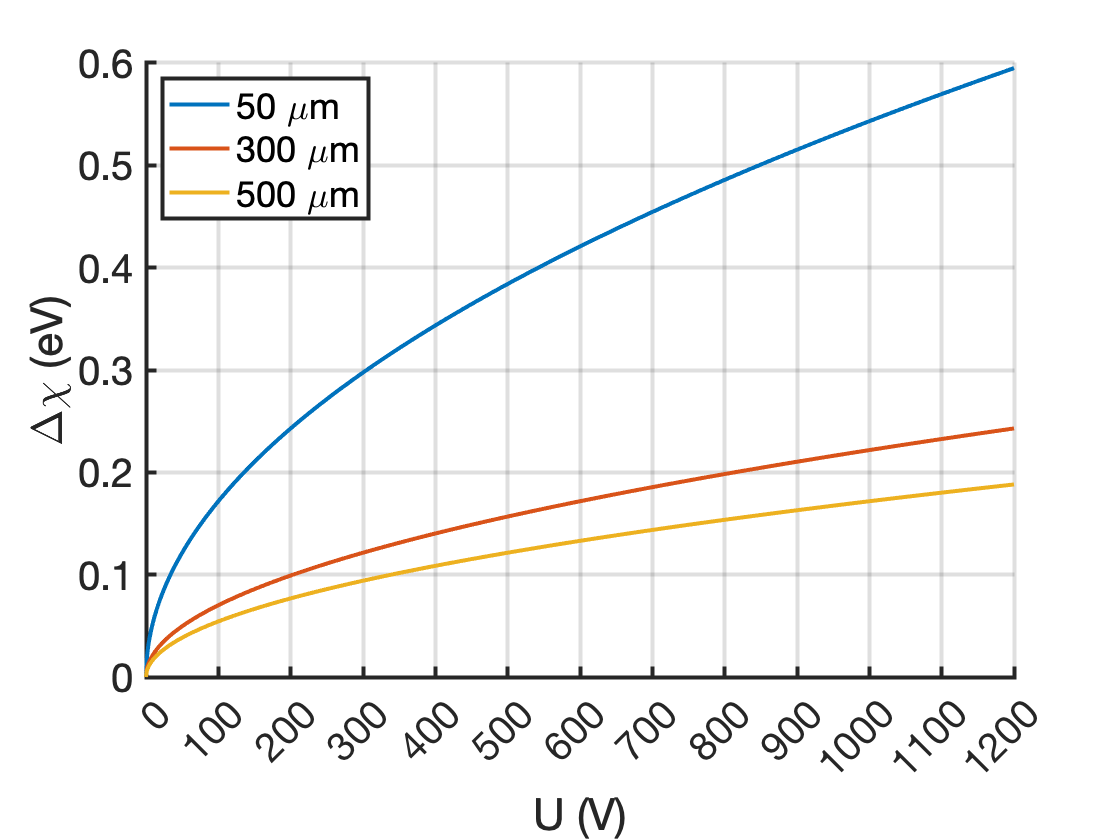}
\caption{\label{fig:12} Schottky effect on dielectrics. A dielectric constant $\epsilon_r$ of 9.8 is used for MgO. In the measurement setup, the distance $d$ between the emission surface of the tynode and the collector is $\SI{50}{\um}$ and the bias voltage is 50 - \SI{200}{\V}.  In TiPC, the distance $d$ between two tynodes is approximately $\SI{500}{\um}$ and the operating voltage is \SI{1000}{\V}. The electric field is in the same order of magnitude in both cases.}
\end{figure}

The proximity of the emission surface of a tynode to the collector seems to improve the TEY. The electric field is estimated to be \SI{2e6}{\V/\m}, which can be increased by using a higher bias voltage between the sample and collector. In figure \ref{fig:11a}, the bias voltage is increased to 100 and \SI{200}{\V}. There is a small improvement in TEY in both cases. However, the effect seems to level off. A similar effect was observed in figure \ref{fig:11b} for the membrane with $d=\SI{15}{\nm}$. A strong electric field near the emission surface of a tynode can have several effects on the (secondary) electron emission: Schottky barrier lowering, field-enhanced electron emission or field emission. The latter two effects require relatively large electric fields and are unlikely to occur in this experiment. Field-enhanced secondary electron emission is observed for electric field strength between $\SI{6.6e7}{\V/\m}$ and $\SI{2.5e8}{\V/\m}$; a PE is needed to initiate SE emission, but the number of SEs is significantly larger than unassisted SE emission \cite{Qin2011}. Above $\SI{2.5e8}{\V/\m}$, field emission occurs at which electrons are extracted without the need of PEs. The Schottky effect is known to lower the work function of cathodes due to the presence of a strong electric field. For dielectrics, Schottky barrier lowering is given by:

\begin{equation}
\label{eq:4}
\Delta \chi = -q\sqrt{\frac{qKF_e}{\epsilon_0}} \quad \text{with} \quad K=\frac{\epsilon_r+1}{\epsilon_r-1}
\end{equation}

where $q$ is the charge in \si{C}, $F_e$ the electric field in \si{V/m}, $\epsilon_0$ the permittivity in vacuum and $\epsilon_r$ the dielectric constant \cite{Cazaux1999}. In figure \ref{fig:12}, the change in the electron affinity is shown as a function of the bias voltage. The electric field depends on the distance and bias voltage: $F_e=U/d$. In this work, the gap is \SI{50}{\um} between the sample and collector. In TiPC, the distance between two tynodes is approximately \SI{500}{\um}, albeit operated with a higher bias voltage. The electric field in both cases is in the same order of magnitude. The effect of electron affinity lowering on secondary electron emission have been estimated by Cazaux \cite{Cazaux2011}. He attributed the discrepancies in (reflection) secondary electron yield data of common materials, such as silicon and aluminum, to differences in work function or electron affinity of the samples due to oxidation and/or contamination. The relative change in the escape probability of SEs due to a change in electron affinity is given by

\begin{equation}
\label{eq:5}
A \sim A_1(\chi/\chi_1)^{-p}
\end{equation}

where $A_1$ is the initial escape probability, $\chi_1$ the initial electron affinity, $\chi$ the altered electron affinity and $p$ a material dependent constant. An increase in the escape probability of SEs will result in a proportional increase in the SE yield. Unfortunately, $p$ have not been reported for magnesium oxide, so the relative change in TEY cannot be estimated and compared to our results. It does show that the escape probability of SEs increases exponentially as the electron affinity is being lowered. However, this contradicts with our results as the effect seems to level off when the bias voltage is increased from 100 to \SI{200}{\V}.

\section{Conclusion \& Outlook}
\label{sec:conclusion}

We have developed a new surface scan method that requires a lower electron dose, which minimizes charge-up and/or contamination effects. In addition, the method allows us to investigate multiple membranes on a tynode. We have also demonstrated that ALD MgO is a viable membrane material for tynodes. Membranes that consists of TiN/MgO layers with layer thicknesses of 2 and $\SI{5}{\nm}$, respectively, provide an averaged maximum TEY of $4.6\pm0.2$ (\SI{1.35}{\keV}). If we limit the operating voltage between two tynodes to \SI{1}{\kV}, then a TEY of 3.7 is still achievable. With a TEY of 3.7, a stack of 6 tynodes would be sufficient to trigger the CMOS chip and allow for single photon detection.

The strong electric field between the tynodes in a stack is beneficial to the TEY due to the Schottky effect, which lowers the electron affinity on the emission surface. The averaged maximum TEY improved from $4.6\pm0.2$ to $5.0\pm0.3$. In future designs, there is room to benefit more from this effect by either reducing the substrate thickness and/or applying a higher bias voltage between the tynodes. Also, an outlier with a TEY of 5.5 was measured on one individual membrane. Thus, improving the fabrication process to ideal conditions can increase the averaged maximum TEY of a tynode.  

The variance in TEYs of the membranes across the array can be reduced by incorporating the TiN deposition earlier in the fabrication process rather than as a post-process at the end. In ref \cite{Chan2021}, we have demonstrated that the TiN layer can be encapsulated by two layers of ALD Al\textsubscript{2}O\textsubscript{3}, which were the electron emission layers in that design. The surface on which the TiN was sputtered was relatively flat, which gave a more uniform layer in comparison with a post-process deposition. As an alternative, titanium nitride can be deposited by ALD, which is even more uniform and allows for a precise controlled deposition rate. We have demonstrated that ALD TiN can be applied as the conduction layer in corrugated films \cite{Chan2022}. 

The ageing mechanism of the MgO membranes needs to be investigated further. It is not yet clear what the mechanisms are that lower the TEY. The tynodes are exposed and kept in ambient conditions, which can change their composition over time. Also, electron-induced contamination was observed after prolonged irradiation. Literature on electron induced contamination has shown that the vacuum level determines the deposition rate of contaminants. The SEM chamber that is used for the measurements presented in this work has a vacuum level of 10\textsuperscript{-5} mbar, which is not ideal. The TiPC detector will be operated at ultra-high vacuum levels (10\textsuperscript{-9} mbar) at which electron induced contamination is minimized. 

The next step in the development of TiPC is the assemblage of a tynode stack with 5 or 6 tynodes. There are two technical challenges that need to be addressed: alignment and SE focusing. First, the arrays of each stage need to be aligned, so that an amplification 'channel' is formed above each pixel pad of a TimePix chip. Alignment grooves are proposed to self-align multiple tynodes \cite{Prodanovic2019}. Second, some SE focusing is required to ensure that all transmission SEs are directed to the active membrane surface of the next amplification stage. Dome-shaped tynodes have been proposed as a solution. Also, corrugated membranes have been fabricated that have a near 100\% active surface \cite{Chan2022}. The 3D-structure of the membrane will have a focusing effect as well. The fabrication process of both designs can be adapted to replace ALD Al\textsubscript{2}O\textsubscript{3} by ALD MgO.

Once the tynode stack is assembled, it can be mounted in a prototype TiPC: the TyTest setup \cite{VanDerReep2020}. The setup consists of an electron gun as electron source, a mount to hold a tynode (stack) and a TimePix Chip as read-out. The vacuum chamber can be modified to allow for more high voltage connections, which are needed for a multi-layer stack. Also, the chamber operates at a higher vacuum level in the order of \SI{1e-8}{\milli\bar}, which can prevent contamination that was observed in this work. The setup was used as an alternative method to determine the TEY of a (single) tynode using a TimePix chip. The results were in agreement with the results obtained with a SEM-based method \cite{Chan2022}.

\acknowledgments
This work is supported by the ERC-Advanced Grant 2012 MEMBrane 320764. We are grateful to A.U. Mane and J.W. Ellam from Argonne National Laboratory for their research on ALD MgO and for depositing ALD MgO on our samples. Many thanks to Else Kooi Lab for providing the training and facilities to manufacture the tynodes. Lastly, we would like to thank H. Akthar, T. ten Bruggencate, W.J. Landgraaf, D. Rotman, B.Looman, T. v.d. Reep,  and C. Hansson for their contributions to the MEMBrane project. 

\printglossary[title={List of Abbreviations}] 

\bibliographystyle{JHEP}
\bibliography{main.bbl}



\end{document}